\documentclass[preprint,showpacs,preprintnumbers,amsmath,amssymb]{revtex4}

\usepackage{graphicx}
\usepackage{dcolumn}
\usepackage{bm}

\begin{document}


\title{Dark Matter and Gauge Coupling Unification in A Supersymmetry 
Model with Vector-like Matter}

\author{Chun Liu and Jia-Shu Lu}
\affiliation{State Key Laboratory of Theoretical Physics 
and Kavli Institute for Theoretical Physics China, \\
Institute of Theoretical Physics,
Chinese Academy of Sciences,\\
P.O. Box 2735, Beijing 100190, China}
 \email{liuc@mail.itp.ac.cn, lujiashu@itp.ac.cn}

\date{\today}

\begin{abstract}
WIMP dark matter and gauge coupling unification are considered in an 
R-parity violating MSSM with vector-like matter.  Dark matter is 
contained in an additional vector-like SU(2)$_L$ doublet which possesses 
a new U(1) gauge symmetry.  The Higgs fields are extended to be in a 
${\bf 5\oplus \bar{5}}$ representation of SU(5).  The stability of dark matter is a result of gauge symmetries, and the mass of the dark matter 
particle is between (1.1-1.5) TeV.  Dark matter has a 
very small cross section with nucleis, thus the model is consistent 
with current dark matter direct detection experiments such as Xenon100.  
The model also predicts new charged and colored particles to be observed 
at LHC.  
\end{abstract}



\maketitle

\section{Introduction}

Dark matter is one of the most important problems in elementary physics 
\cite{review1}.  It plays a very important role in understanding some 
astro-physical observation.  If dark matter particles are thermally 
produced in the early Universe, an attractive scenario appears, namely 
that the dark matter particle has a hundreds GeV mass with typically 
weak interaction \cite{review2}.  The weakly interacting massive 
particle (WIMP) scenario of dark matter is interesting for particle 
physics.  In high energy physics experiments searching for the 
electro-weak symmetry breaking (EWSB) mechanism, WIMP should be found in 
the near future.  

The particle physics beyond the Standard Model (SM) has the following 
main stream logic.  SM gauge interactions unify at a high scale 
($\sim 10^{16}$) GeV \cite{gut}.  Supersymmetry (SUSY) \cite{susy} is 
then required to stabilize the Higgs mass.  The minimal SUSY extension 
of SM (MSSM) which necessarily involves two Higgs doublets makes the 
idea of the grand unification theory (GUT) more meaningful due to LEP 
data \cite{lep}.  WIMP dark matter is right the lightest neutralino 
\cite{review2} by further assuming R-parity conservation, . 

We will work in the SUSY paradigm without assuming R-parity 
conservation.  
R-parity conservation is usually adopted to avoid rapid decays of the 
proton, however, it lacks of motivation from first principles.  Instead, 
we can only assume baryon number conservation which is also 
phenomenologically viable.  R-parity violation makes things more 
complicated.  The lightest neutralino is no longer stable.  To have  
dark matter, new particles are needed then.  To keep dark matter 
WIMPs, the simplest realization is to introduce a vector-like SU(2)$_L$ 
doublet, like the two Higgs doublets in MSSM, with a weak scale mass.  
We take one fermionic neutral component of the new doublet superfields 
as the dark matter particle.  Its stability requires a new symmetry 
which is a gauge symmetry.  The necessity of a new gauge symmetry, 
instead of a discrete symmetry, lies in the following point.  The 
elastic scattering of the dark matter particle via a $Z$ boson exchange 
should be suppressed.  In our case, the dark matter particle is not 
really a Dirac one, because spontaneous breaking of the new gauge 
symmetry splits the neutral particle spectrum.  As a result, dark 
matter and the $Z$ boson interaction is almost inelastic.  

Now we consider how to let the new dark matter particle content 
compatible with gauge coupling unification.  Because one vector-like 
doublet is introduced, in addition to the particles of MSSM, gauge 
coupling constants would no longer unify in this case.  GUT relation of 
gauge coupling constants could be restored if some colored particles 
are further added so that all the new particles form complete 
representations of GUT \cite{newdm,1}.  Alternatively, we can keep the 
dark matter sector as simple as possible, namely it is just SUSY 
two-Higgs-doublet alike, but we attach the new colored particles with 
the two Higgs doublets.  In an effort of extending MSSM, we proposed 
that Higgses are understood as sleptons of an extra vector-like 
generation \cite{liu}.  In that model, GUT was lost, and there was no 
dark matter candidates.  It is interesting to note that whence we 
consider the above discussed dark matter scenario, GUT relation of the 
gauge coupling constants can be restored.  This will make the whole low 
energy SUSY model more meaningful.  

However, careful consideration about running gauge coupling constants 
tells us that things are not that easy and straightforward.  The 
particle content of the extra vector-like generation \cite{liu} needs to 
be reduced.  To avoid Landau poles of the coupling constants, new 
particles at the TeV scale cannot be that many.  Instead of a TeV 
vector-like generation which would have both a ${\bf 5\oplus \bar{5}}$ 
and a ${\bf 10\oplus \bar{10}}$ representations, we are only allowed to 
have a TeV ${\bf 5\oplus \bar{5}}$ in which the two Higgs doublets are 
contained.  Therefore, as far as the Higgs content is concerned, we 
return back to the ordinary GUT.  Here the new point is that the 
$\bf{5}$ representation will mix with ordinary three generation 
fermions, and the triplet Higgs mass is around TeV.  We impose by hand 
at the moment baryon number conservation.  

In the next section, we present the model which includes the dark matter 
sector.  In Sect. III, dark matter properties and collider phenomenology 
are analyzed.  Discussions and the summary are given in the last two 
sections.

\section{Model}

Let us first look at the SM relevant sector.  
Within the framework of SUSY 
and SU(3)$_c\times$SU(2)$_L\times$U(1)$_Y$ gauge symmetry, the particle 
content is extended in a way that the two Higgs doublets are contained 
in a SU(5) ${\bf 5\oplus \bar{5}}$ representation.  These vector-like 
particles have masses of (100-1000) GeV.  Because of R-parity violation 
there is no dark matter particle.  

To include the dark matter sector, we note that all the particles 
already introduced are in full representations of SU(5) GUT.  For the 
GUT purpose, the WIMP sector should be exactly composed of a pair of 
SU(2)$_L$ doublets with opposite U(1)$_Y$ charges.  This is the reason 
to take them SUSY two-Higgs-doublets-like.  To avoid extra degrees of 
freedom which may violate GUT relation, the new interaction among the 
WIMPs should be another Abelian gauge interaction U(1)$_n$.  
The U(1)$_n$ charge can be arranged so that after U(1)$_n$ breaking, 
an unbroken Z$_2$ symmetry will remain. 
This Z$_2$ symmetry makes dark matter stable.  

The model is SUSY SU(2)$_L\times$U(1)$_Y\times$SU(3)$_c\times$U(1)$_n$ 
gauge symmetric one with baryon number conservation.  The particle 
content is given below with their quantum numbers under above gauge 
symmetries and global baryon numbers, 
\begin{equation}
\label{1}
\begin{array}{c}
L_m(2,-1,1,0,0)\,, E_i^c(1,2,1,0,0)\,,Q_i(2,\frac{1}{3},3,0,\frac{1}{3})
\,, U_i^c(1,-\frac{4}{3},\bar{3},0,-\frac{1}{3})\,, 
D_m^c(1, \frac{2}{3},\bar{3},0,-\frac{1}{3})\,, \\[3mm]
H_u(2,1,1,0,0)\,, D_H^c(1,-\frac{2}{3}, 3, 0, \frac{1}{3})\,, 
\end{array}
\end{equation} 
and 
\begin{equation}
\label{2}
\chi_1(2,-1,1,-1,0)\,, \chi_2(2,1,1,1,0)\,, \phi_1(1,0,1,-2,0)\,, 
\phi_2(1,0,1,2,0)\,, X(1,0,1,0,0,0)\,.  
\end{equation} 
Field notation is conventional, like $L$ stands for lepton doublets, 
$E^c$ for lepton singlets, $Q$ for quark doublets, $U^c$ - $D^c$ for 
quark singlets and $H_u$ for the up-type Higgs doublet.  
In Eq. (\ref{1}), $m=1-4$ and $i=1-3$.  One 
combination of $L_m$'s makes the down-type Higgs.  In addition to the 
ordinary three generations, there is a vector-like 
${\bf 5}$-representation ($L_4$, $D_4^c$ and $H_u$, $D_H^c$).  The dark 
matter sector is given in Eq. (\ref{2}).  Two doublets $\chi_{1,2}$ 
contain the WIMP, and $\phi_{1,2}$ are the U(1)$_n$ breaking Higgs.

\subsection{SM relevant part}  

The superpotential of our sector containing the SM part can be written 
down.  Instead of R-parity, baryon number conservation is assumed, 
\begin{equation}
\label{01}
{\mathcal W}= \mu_m L_m H_u + \mu_m^D D_m^c D_H^c
              +\lambda_{mni} L_m L_n E_i^c + \lambda'_{imn} Q_i L_m D_n^c 
              + y_{ij} Q_i H_u U_j^c + \tilde{y}_{ij} E_i^c D_H^c U_j^c \,,
\end{equation} 
where $\mu_m$'s are mass parameters, $\lambda^{(\prime)}$, $y$ and 
$\tilde{y}$'s coefficients.  By redefining the down-type Higgs and the 
fourth down-quark field, 
\begin{equation}
\label{02}
H_d\equiv \frac{\mu_m}{\mu}L_m\,,~~
D_4^c\equiv\frac{\mu_m^D}{\mu^D}D_m^c\,,~~
\end{equation}
where 
\begin{equation}
\label{03}
\mu\equiv \sqrt{\sum_{m=1}^{4}|\mu_m|^2}\,,~~
\mu^D\equiv \sqrt{\sum_{m=1}^{4}|\mu_m^D|^2}\,,
\end{equation} 
the SM relevant superpotential becomes to be  
\begin{equation}
\label{04}
\begin{array}{lll}
{\mathcal W}&=& \mu H_d H_u+ \mu^D D_4^c D_H^c+y_{ij}^l L_i H_d E_j^c 
                + y_{ij}^d Q_i H_d D_j^c + y_{ij} Q_i H_u U_j^c \\
            & & +\lambda_{ijk}L_i L_j E_k^c 
                +\lambda'_{ijk}Q_i L_jD_k^c+\lambda_{ij}^DQ_iL_jD_4^c 
                + y_i^D Q_i H_d D_4^c 
                + \tilde{y}_{ij} E_i^c D_H^c U_j^c \,,
\end{array} 
\end{equation}
where field decomposition have been generally written as follows, 
\begin{equation}
\label{05}
L_m   = c_{mi} L_i     + c_{m4} H_d     \,, ~~ 
D_m^c = c_{mi}^D D_i^c + c_{m4}^D D_4^c \,, 
\end{equation} 
and the coefficients are 
\begin{equation}
\label{06}
\begin{array}{c}
y_{ij}^l = 2\lambda_{mnj} c_{mi}c_{n4} \,,~~ 
y_{ij}^d = \lambda'_{imn} c_{m4}c_{nj}^D \,,~~
\lambda_{ijk}  = \lambda_{mnk} c_{mi}c_{nj} \,,\\
\lambda'_{ijk} = \lambda'_{imn} c_{mj}c_{nk}^D \,,~~ 
\lambda_{ij}^D = \lambda'_{imn} c_{mj}c_{n4}^D \,,~~ 
y_i^D          = \lambda'_{imn} c_{m4}c_{n4}^D \,. 
\end{array}
\end{equation} 

From the superpotential (\ref{04}), we see that because of Dirac mass 
terms of up-type Higgs and the four doublet leptons, $D_H^c$ and the 
four singlet down-quarks, one of the four lepton doublets and one of the 
down-quarks, namely the fourth doublet lepton $H_d$ and the fourth 
singlet down-quark $D_4^c$ are always heavy, $H_d$ is identified as the 
down-type Higgs.  The fourth neutrino together with the "neutrino" in 
$H_u$ consists of 
neutral Higgsinos.  After the mass terms, the next five terms 
in Eq. (\ref{04}) are ordinary Yukawa interactions and trilinear lepton 
number (R-parity) violating terms.  The other three terms in (\ref{04}) 
are new which involve the $D_{4(H)}^c$ field.  
Two of them also violate lepton 
numbers.  

Soft SUSY breaking mass terms should be included into the Lagrangian.  
In addition to gaugino masses, they include mass-squared terms of 
scalars and $B\mu$-type terms corresponding to those $\mu$-terms in 
superpotential (\ref{01}), 
\begin{equation}
\label{07} 
\begin{array}{lll}
-{\mathcal L}&\supset&
                   M^2\tilde{L}_m^{\dag}\tilde{L}_m + M_h^2h_u^{\dag}h_u 
                      + M_E^2\tilde{E}_i^{c\dag}\tilde{E}_i^c 
                      + M_Q^2\tilde{Q}_i^{\dag}\tilde{Q}_i 
                      + M_U^2\tilde{U}_i^{c\dag}\tilde{U}_i^c 
                      + M_D^2\tilde{D}_m^{c\dag}\tilde{D}_m^c \\
            &        &+ M_{DH}^2\tilde{D}_H^{c*}\tilde{D}_H^c 
                      +(B\mu_m\tilde{L}_mh_u
                      +B^D\mu_m^D\tilde{D}_m^c\tilde{D}_H^c + h.c.)\,,
\end{array}
\end{equation}
where tildes stand for scalars.  We have assumed universality of the 
mass-squared terms and the alignment of the $B$ terms, namely both the 
mass parameters $B$ and $B^D$ do not depend on the sub-script $m$.  In 
terms of three light generations of Eq. (\ref{04}), universality of 
these soft mass terms is easily seen, 
\begin{equation}
\label{08} 
\begin{array}{lll}
-{\mathcal L}&\supset& M^2\tilde{L}_i^{\dag}\tilde{L}_i+M^2h_d^{\dag}h_d  
            + M_h^2h_u^{\dag}h_u + M_E^2\tilde{E}_i^{c\dag}\tilde{E}_i^c 
                      + M_Q^2\tilde{Q}_i^{\dag}\tilde{Q}_i 
                      + M_U^2\tilde{U}_i^{c\dag}\tilde{U}_i^c 
                      + M_D^2\tilde{D}_m^{c\dag}\tilde{D}_m^c \\
             &       &+ M_{DH}^2\tilde{D}_H^{c*}\tilde{D}_H^c 
                      +(B\mu h_dh_u+B^D\mu^D\tilde{D}_4^c\tilde{D}_H^c 
                      + h.c.)\,.
\end{array}
\end{equation}
Numerically soft masses $M$'s, $B$'s and gaugino masses are assumed to 
be $\cal O$(100) GeV.  
Soft trilinear terms corresponding to Eq. (\ref{01}) are 
\begin{equation}
\label{09}
{\mathcal L} \supset
             \bar{\lambda}_{mni} \tilde{L}_m \tilde{L}_n \tilde{E}_i^c 
             +\bar{\lambda}'_{imn} \tilde{Q}_i\tilde{L}_m\tilde{D}_n^c 
             +\bar{y}_{ij} \tilde{Q}_i h_u \tilde{U}_j^c 
             +\bar{\tilde{y}}_{ij}\tilde{E}_i^c\tilde{D}_H^c\tilde{U}_j^c 
             + h.c. \,, 
\end{equation} 
where the following coupling alignment will be assumed, 
\begin{equation} 
\label{010} 
\bar{\lambda}_{mni} = \lambda_{mni} m_0\,,~~
\bar{\lambda}'_{imn} = \lambda'_{imn} m_0\,,~~
\bar{y}_{ij} = y_{ij}m_0 
\end{equation}
with $m_0$ being the order of soft masses $\sim{\cal O}(100)$ GeV.  

Let us look at gauge symmetry breaking.  
From the Lagrangian, the scalar potential can be written down 
straightforwardly.  To get EWSB, one needs a negative determinant of 
the Higgs mass-squared matrix, namely 
\begin{equation}
\label{010a}
\left(M^2  +\mu^2\right) \left(M_h^2+\mu^2\right) < |B\mu|^2\,
\end{equation}
with the ordinary condition $M^2+M_h^2+2\mu^2+2B\mu>0$.  This 
requirement can be realized when the renormalization group is considered.  
$M_h^2$ will become negative at the weak scale, due to the large top 
quark Yukawa coupling.  Therefore, everything of EWSB here will be the 
same as that in MSSM.  The MSSM analysis of EWSB applies here.  EWSB 
in this model occurs at the weak scale.  Besides Eq. (\ref{010a}), 
correct EWSB also requires 
\begin{equation}
\label{013} 
\left(M_D^2+\mu^{D2}\right)\left(M_{DH}^2+\mu^{D2}\right) > 
|B^D\mu^D|^2 \,.   
\end{equation}
Then the remaining analysis of EWSB is identical to that of MSSM with 
same Higgs and Higgsino spectra.  Eq. (\ref{013}) can be satisfied 
easily.  Carefully thinking of EWSB conditions Eqs. (\ref{010a}) and 
(\ref{013}), we see that if $\mu<\mu^D$, EWSB occurs naturally.  

\subsection{Dark sector}  

The dark matter sector Lagrangian is written according to the gauge 
invariance, 
\begin{equation}
\label{3}
\begin{array}{lll}
{\mathcal L}_{\rm dark}&=&\left(\chi_1^\dag e^{g_2V_2+g_1V_1+g_1'V'_1}
\chi_1+ \chi_2^\dag e^{-g_2V_2-g_1V_1-g_1'V'_1}\chi_2 
+ \phi_1^\dag e^{2g_1'V'_1}\phi_1 + \phi_2^\dag e^{-2g_1'V'_1}\phi_2
+ X^{\dag}X
\right)\left|_{\theta\theta\bar{\theta}\bar{\theta}}\right.\\[3mm]
& & + \left(\mu' \chi_1 \chi_2\left|_{\theta\theta}\right. 
+ cX(\phi_1\phi_2-\mu^{\prime\prime2})\left|_{\theta\theta}\right. 
+ h.c. \right)\,.
\end{array}  
\end{equation} 
where $\mu'$ and $\mu''$ are mass parameters, $g_1'$ and $c$ coupling 
constants.  It is important to note that an accidental Z$_2$ discrete 
symmetry appears, under which $\chi_1$ and $\chi_2$ fields are odd and 
all the other fields are even.  As we will see, it remains unbroken even 
after U(1)$_n$ breaking as well as EWSB.  The Z$_2$ symmetry keeps the 
lightest component of $\chi_i$ $(i=1,2)$ stable.  That is the dark 
matter particle in this model.  

When SUSY breaking is considered, soft masses should be included, 
\begin{equation}
\begin{array}{lll}
\label{4}
{\mathcal L}_{\rm dark, soft}&=&
-\frac{1}{2}m_1'\lambda^{1\prime}\lambda^{1\prime}
+ m_{\tilde{\chi}_1}^2 \tilde{\chi}_1^* \tilde{\chi}_1 
+ m_{\tilde{\chi}_2}^2 \tilde{\chi}_2^* \tilde{\chi}_2 
+ m_{\phi_1}^2 \phi_1^* \phi_1 + m_{\phi_2}^2 \phi_2^* \phi_2 \\
& & + m_x^2 x^* x + (B'\mu')\tilde{\chi}_1 \tilde{\chi}_2 + h.c. ) \,. 
\end{array}
\end{equation}
U(1)$_n$ gauge symmetry breaks spontaneously when $\phi_{1,2}$  get 
non-vanishing vacuum expectation values (VEVs).  From Eqs. (\ref{3}) 
and (\ref{4}), the relevant scalar potential for $\langle\phi_1\rangle$ 
and $\langle\phi_2\rangle$ determination is 
\begin{equation}
\label{5}
V_{\rm dark} = 2g^{\prime 2}_1\left(\langle\phi_1\rangle^2
-\langle\phi_2\rangle^2\right)^2 
+ c^2|\langle\phi_1\rangle \langle\phi_2\rangle - \mu''^2|^2 
+ m_\phi^2\left(\langle\phi_1\rangle^2+\langle\phi_2\rangle^2 \right)\,,
\end{equation} 
where it has taken $m_\phi^2 = m_{\phi_1}^2 = m_{\phi_2}^2$.  Then 
\begin{equation}
\label{6}
\langle\phi_1\rangle = \langle\phi_2\rangle = \langle\phi\rangle =
\left(\mu^{\prime\prime2} - \frac{m_\phi^2}{c} \right)^{\frac{1}{2}} \,. 
\end{equation} 
Vector-like particle masses are all taken to be similar \cite{liu}, 
hence $\mu', ~ \mu''\sim 1$ TeV.  It is natural to expect that  
$\langle\phi\rangle\sim \cal O$(100) GeV.  The U(1)$_n$ gauge boson 
$\gamma_n$ gets a mass of $m_{\gamma_n}=4g_1'\langle\phi\rangle$.  
As long as the bosonic fields of SU(2)$_L$ doublets $\chi_{1,2}$ do not 
get VEVs and are heavy enough, the Z$_2$ symmetry still remains after 
U(1)$_n$ breaking. 

In addition to  Eqs. (\ref{3}) and (\ref{4}), the Lagrangian should 
include a gauge field mixing between U(1)$_n$ and U(1)$_Y$, 
\begin{equation}
{\mathcal L}_{\rm mixing} = \epsilon F_n^{\mu\nu} F_{Y\mu\nu}\,.  
\end{equation} 
This mixing is accompanied with a gaugino mixing because of SUSY,
\begin{equation}
{\mathcal L}_{\rm gaugino ~mixing} = 
2\epsilon (\lambda^1\sigma^{\mu}\partial_{\mu}\bar{\lambda}^1{'} 
+ \lambda^1{'}\sigma^{\mu}\partial_{\mu}\bar{\lambda}^1)\,.  
\end{equation} 
It is conventional to choose the mixing to be of the size of 
$\epsilon \sim 10^{-3}$.  This mixing makes possibly lighter particles 
in the dark sector, such as $\phi_{1,2}$, decay into MSSM particles.  

What we are interested in is the spectrum of $\chi_1$ and $\chi_2$ 
particles, because they carry SM quantum numbers.  For the fermions, at 
the leading order, they form a Dirac particle 
$\Psi_\chi=\left(\begin{array}{c}\chi_1\\ \bar{\chi}_2\end{array}\right)$
with a mass $\mu'$, 
\begin{equation}
\label{7}
{\mathcal L}_\chi = \bar{\Psi}_{\chi} i\gamma_\mu D^\mu \Psi_{\chi} 
- \mu' \bar{\Psi}_{\chi} \Psi_{\chi} \; , 
\end{equation} 
where 
$D_\mu=\partial_\mu-ig_2A_\mu^a\tau^a-ig_1A_\mu^1-ig_1'A^{\prime 1}_\mu$.  
Actually $\Psi_\chi$ is a pseudo-Dirac particle because of gauge 
symmetry breaking.  Generally, 
EWSB splits the neutral and charged components of $\Psi_\chi$, 
and U(1)$_n$ breaking further splits the two neutral components. 
In this model, such mass splittings are described by the following gauge 
symmetric dimension 5 and 6 operators, 
\begin{equation}
{\mathcal L}\supset {\mathcal L}^{\rm dim.5}+{\mathcal L}^{\rm dim.6}\,,  
\end{equation}
where 
\begin{equation}
\label{8}
\begin{array}{lll} 
{\mathcal L}^{\rm dim.5} & = & \displaystyle \frac{a_1}{\Lambda}
(\chi_1 H_u)(\chi_2 H_d)\left|_{\theta\theta}\right. 
+\frac{a_2}{\Lambda}(\chi_1 H_d)(\chi_2 H_u)\left|_{\theta\theta}\right.
+ h.c.\,,\\[5mm]
{\mathcal L}^{\rm dim.6} & = & \displaystyle  
 \frac{a_3}{\Lambda^2}\phi_2(\chi_1 H_u)
(\chi_1 H_u)\left|_{\theta\theta}\right. 
+\frac{a_4}{\Lambda^2}\phi_1(\chi_2 H_d)
(\chi_2 H_d)\left|_{\theta\theta}\right.
+ h.c.\, \\[3mm]
\end{array}
\end{equation}
with $\Lambda$ being a cutoff, and $a_i \sim \cal O$(1) coefficients.  
$\Lambda$ may be considered as the scale of SUSY breaking messengers 
which have been integrated out, the messengers also form complete SU(5) 
representations and do not break the unification of SM gauge couplings.  
As it will be seen, $\cal{L}^{\rm dim.5}$ splits charged and neutral 
components, and $\cal{L}^{\rm dim.6}$ splits the two neutral ones.   

Note that without dimension-5 operators, EWSB itself at the 
renormalizable level induces mass 
splitting between charged and neutral components.  One-loop diagrams 
with a Z boson propagating in inner lines directly give splitting 
roughly as $\frac{\alpha_2}{4\pi}M_Z$.  It is of the order 0.1 GeV.  For 
splitting the two neutral parts, we nevertheless need the new higher 
dimensional operators of $\cal{L}^{\rm dim.6}$.  In such a situation, it 
is natural to expect that a general $\cal{L}^{\rm dim.5}$ is also there.  
In our analysis $\cal{L}^{\rm dim.5}$ is taking as the effective 
operators which parameterize all the EWSB effects.  

$a_1$ and $a_2$ terms simply give rise to mass splitting between the 
charged and the neutral components, 
\begin{equation}
\Delta M = (a_1+a_2) \frac{v^2 \sin 2\beta}{4\Lambda} \,,  
\end{equation}
where $v=246$ GeV.  With tan$\beta$ between $(3-10)$ and $\Lambda$ 
between $(10-100)$ TeV, this splitting ranges from $(0.1-1)$ GeV.  
Here the positive 
$(a_1+a_2)$ case is chosen so that the neutral components of $\chi_i$, 
$\chi_1^0$ and $\chi_2^0$, are lighter.  The charged components 
$\chi_1^-$ and $\chi_2^+$ form an exact Dirac particle.  We can see that 
mass splitting generated by dimension 5 operators is at least as large 
as that generated by loop diagrams.  

$\chi_1^0$ and $\chi_2^0$ are splitted further through the 
$a_3$ and $a_4$ terms.  The mass matrix of $\chi_1^0 , \chi_2^0$ turns 
out to be
\begin{equation}
\label{13}
\mathcal M = \left(
\begin{array}{cc}
\displaystyle (\frac{v}{\Lambda})^2 
\frac{\langle\phi\rangle}{2} a_3\sin^2\beta & 
\displaystyle 
\frac{1}{2}\left[\mu'-\frac{v^2 \sin 2\beta}{4\Lambda}(a_1 + a_2)
\right] \\ [3mm] 
\displaystyle \frac{1}{2}\left[\mu'-\frac{v^2 \sin 2\beta}{4\Lambda}
(a_1 + a_2)\right] & 
\displaystyle 
(\frac{v}{\Lambda})^2 \frac{\langle\phi\rangle}{2} a_4{\rm cos}^2\beta \\
\end{array}
\right)\,.\\[3mm]
\end{equation}
As the off-diagonal elements are much larger than the diagonal ones, 
the mass eigenvalues are approximately 
$\displaystyle \mu'-\frac{v^2\sin2\beta}{4\Lambda}(a_1+a_2)\pm
(\frac{v}{\Lambda})^2\frac{\langle\phi\rangle}{2}
(a_3\sin^2\beta+a_4\cos^2\beta)$, 
and the corresponding mass eigenstates are approximately 
\begin{equation}
\label{14}
\chi'_d = (\chi_1^0 + \chi_2^0 )/\sqrt{2} 
+ \mathcal{O}(\frac{v^2\langle\phi\rangle}{\Lambda^2\mu'}) \,, ~~~ 
\chi_d = i(\chi_1^0 - \chi_2^0 )/\sqrt{2} 
+ \mathcal{O}(\frac{v^2\langle\phi\rangle}{\Lambda^2\mu'}) \\[3mm]
\end{equation}
with $\chi_d$ being the lighter state.  Therefore, mass splitting 
between the two neutral Majorana fermions is
\begin{equation}
\Delta m=\left(\frac{v}{\Lambda}\right)^2\langle\phi\rangle
(a_3\sin^2\beta+a_4\cos^2\beta)\,.\\[3mm]  
\end{equation}
This splitting is almost independent of $\tan\beta$.  Taking 
$\langle\phi\rangle\sim 100$ GeV, 
this splitting ranges from $(1-100)$ MeV.

The other particles of the dark sector have the following spectrum.  
As $\phi_1, \phi_2$ get VEVs, their fermionic partners 
$\tilde{\phi}_{1,2},\tilde{X}$ which is the fermion of $X$, and gaugino 
$\lambda^{1\prime}$ will get masses.  It is convenient to change the 
basis to
\begin{equation}
\label{16}
\Phi'=(\tilde{\phi}_1 + \tilde{\phi}_2 )/\sqrt{2} \,, ~~~ 
\Phi=(\tilde{\phi}_1 - \tilde{\phi}_2 )/\sqrt{2} \,.  
\end{equation}
$\Phi'$ and $\tilde{X}$ form a Dirac particle with a mass of 
$\sqrt{2} c \langle\phi\rangle \sim 100$ GeV.  
The mass matrix of $\Phi$ and $\lambda^{1\prime}$ is the following by 
further considering the $\lambda^{1\prime}$ soft mass, 
\begin{equation}
\label{18}
\mathcal{M'} = \left(
\begin{array}{cc}
0 & -i4g_1'\langle\phi\rangle\\
-i4g_1'\langle\phi\rangle & m_1'\\
\end{array}
\right)\,.\\[3mm]
\end{equation}
The matrix elements are all $\cal O$(100) GeV, so the mass 
eigenstates $N$ and $N'$ are of the same mass scale, with 
$N$ the lighter one.

The scalars $\tilde{\chi}_{1,2}$ and 
$\phi_{1,2}$ are heavy $\sim 100$ GeV $- 1$ TeV due to soft SUSY 
breaking.  Notice that the singlet $x$ has no gauge couplings and 
can have a vanishing soft mass in the case of gauge mediated SUSY 
breaking.  In this case the boson $x$ will have a mass 
$\sqrt{2} c \langle\phi\rangle\sim 100$ GeV 
which is degenerate to the corresponding fermion.  

It is convenient to express the fermions in the 4-component form:
\begin{equation}
\begin{array}{c}
\Psi_d=\left(\begin{array}{c}\chi_d\\ \bar{\chi}_d\end{array}\right)\,,
\Psi_d'=\left(\begin{array}{c}\chi_d'\\ \bar{\chi}_d'\end{array}\right)\,,
\Psi_-=\left(\begin{array}{c}\chi_{1-}\\ \bar{\chi}_{2+}\end{array}\right)\,.
\end{array}
\end{equation}
In terms of all above mass eigenstates, the dark sector Lagrangian 
relevant to dark matter annihilation can be expressed as 
\begin{equation}
\label{21}
\begin{array}{lll}
{\mathcal L}_{\rm dark}&\supset&
\displaystyle - i\frac{g_2}{2\rm{cos}\theta_W}Z_{\mu}
(\bar{\Psi}_d'\gamma^{\mu}\Psi_d - i\bar{\Psi}_-\gamma^{\mu}\Psi_- 
+ \epsilon'(\bar{\Psi}_d\gamma^{\mu}\Psi_d - \bar{\Psi}_d'\gamma^{\mu}\Psi_d'))
\\[3mm]
& &- ig_1'A_{1\mu}'
(\bar{\Psi}_d'\gamma^{\mu}\Psi_d - i\bar{\Psi}_-\gamma^{\mu}\Psi_- 
+ 2\bar{\Psi}_X\gamma^{\mu}\Psi_N
+ \epsilon'(\bar{\Psi}_d\gamma^{\mu}\Psi_d - \bar{\Psi}_d'\gamma^{\mu}\Psi_d'))
\\[3mm]
& &-g_2{\rm sin}\theta_WA_{\mu}
(\bar{\Psi}_-\gamma^{\mu}\Psi_-)
\\[3mm]
& &\displaystyle +\frac{g_2}{2}W^{+}_{\mu}
(\bar{\Psi}_d'\gamma^{\mu}\Psi_- + i\bar{\Psi}_d\gamma^{\mu}\Psi_-)
\\[3mm]
& &\displaystyle +\frac{g_2}{2}W^{-}_{\mu}
(-\bar{\Psi}_-\gamma^{\mu}\Psi_d' + i\bar{\Psi}_-\gamma^{\mu}\Psi_d)\,.
\end{array} 
\end{equation} 
It is seen that the dark matter particle
$\chi_d$ mainly scatter inelastically via 
gauge interactions.  Note that there are still small diagonal 
$\chi_d$-$\chi_d$-gauge boson couplings which are about  
$\epsilon'\simeq \Delta m / \mu'$ as can be seen from Eq. (\ref{14}).  

\subsection{UV-completion}
Up to now, our model is a TeV effective theory which does not include 
particles much heavier than TeV.  The cut off $\Lambda$ may be, as we 
have mentioned, understood as a result of integrating out SUSY 
breaking messengers in the gauge mediated SUSY breaking scenario.  
Here we present an UV completion model which reproduces our effective 
theory.  

The SUSY breaking messengers have the following quantum numbers of 
SU(2)$_L\times$U(1)$_Y\times$SU(3)$_c\times$U(1)$_n$, 
\begin{equation}
\begin{array}{c}
\eta(2,-1,1,-1)\,, \eta'(1, \frac{2}{3},\bar{3},-1)\,, 
\kappa(1,0,1,-1)\,,\\[3mm]
\bar{\eta}(2,1,1,1)\,, \bar{\eta}'(1,-\frac{2}{3}, 3, 1)\,. 
\bar{\kappa}(1,0,1,1)\,,
\end{array}
\end{equation}
The SUSY breaking spurion field $S$ couples with messengers
\begin{equation}
{\mathcal W}_{spurion} = S(\eta\bar{\eta}+\eta'\bar{\eta}'+\kappa\bar{\kappa})\,, 
\end{equation}
where $S$ get a VEV, 
\begin{equation}
\langle S\rangle = \Lambda + \theta\theta F\,. 
\end{equation}
The soft masses are the same as those of normal gauge mediation. 
For dark sector,
\begin{equation}
\begin{array}{c}
m_{\lambda^{1'}} \sim \displaystyle \frac{\alpha_1'}{4\pi}
\frac{F}{\Lambda}\\[3mm] 
m^2_{\tilde{\chi}_{1,2}} \sim \displaystyle (\frac{\alpha_1'}{4\pi}
\frac{F}{\Lambda})^2\\[3mm] 
m^2_{\phi_{1,2}} \sim \displaystyle (\frac{\alpha_1'}{\pi}\frac{F}{\Lambda})^2 \,.  
\end{array}
\end{equation}

The U(1)$_n$ messengers $\kappa$ and $\bar{\kappa}$ have Yukawa 
couplings with Higgs and dark sector,
\begin{equation}
{\mathcal W}_{messenger} = b_1\chi_1H_u\bar{\kappa} 
+ b_2 \chi_2H_d\kappa 
+ b_3 \phi_2\kappa\kappa + b_4 \phi_1\bar{\kappa}\bar{\kappa}\,.
\end{equation}
These couplings give rise to the $a_i$ couplings on tree level 
by integrating out $\kappa$ and $\bar{\kappa}$,
\begin{equation}
\begin{array}{c}
a_1 = b_1b_2\\[3mm]
a_3 = b_3b_1^2\\[3mm]
a_4 = b_4b_2^2\,.
\end{array}
\end{equation}
Here the lack of $a_2$ will not change mass splitting result
of the dark sector significantly because $a_1$ is non-zero. 
There could be models in which $a_2$ does not vanish.

The Yukawa couplings b$_i$ cause small mixing between 
messengers and dark sector, which makes messengers decay 
into dark sector. After EW and U(1)$_n$ breaking, a Z$_2$ 
symmetry remains in messenger and dark sector, which 
is the same Z$_2$ that appears only in dark sector when 
messengers are integrated out.

\section{Analysis}

\subsection{Dark matter relic density}  

From discussions of the last section and Eq. (\ref{21}), we see that 
the only stable particle in this model is $\chi_d$, which is 
dark matter.  

In relic density calculation, coannihilation of all 
the four components of $\chi_1$ and $\chi_2$ should be considered. 
We use the program micrOMEGAs to calculate relic density
\cite{Belanger:2010pz}\cite{Belanger:2008sj}\cite{Belanger:2006is}.
As long as $\chi_d$ has the correct relic density, the dark matter mass $\mu'$ 
can be determined by the coupling $g_1'$. 
\begin{figure}
\includegraphics{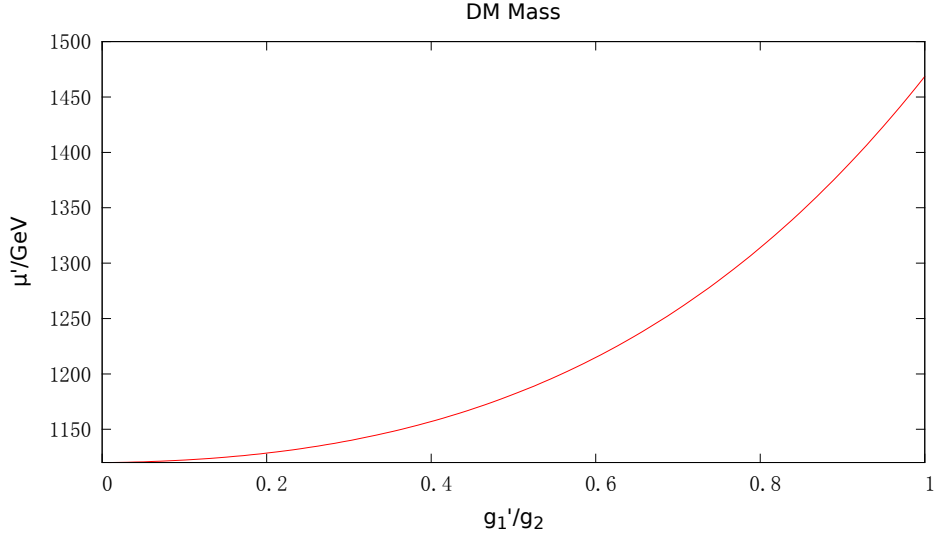}
\caption{\label{fig1}
The relation between dark matter mass $\mu'$ and $g_1'/g_2$.
}
\end{figure}

From FIG \ref{fig1} we can see that the dark matter mass 
is between $\mu'= (1.1\sim1.5)~\rm{TeV}$ when $g_1'$ 
ranges between 0 to $g_2$. This means that taking 
$g_1'\sim g_2$ will not deviate too much
from normal WIMP expectation.  

\subsection{Detection of Dark matter}

Direct detection experiments for dark matter have given strict 
upper limits on cross sections of its scattering with nucleus. As 
$\chi_d$ mainly interacts with Z boson inelastically, it is possible 
to suppress the scattering cross section with nucleus.  
It has been known that if the mass splitting $\Delta m$ between 
$\chi_d$ and $\chi_d'$ is zero, the $\chi_d$-nucleon 
spin-independent cross section would be about $10^{-38}$ cm$^2$ 
\cite{witten} while Xenon100 gives the upper limit of the 
cross section to be $10^{-44}$ cm$^2$ for $\sim$ 1 TeV dark matter 
\cite{xenon}. In our model, $\Delta m$ ranges from (1-100) MeV depending on 
the cutoff $\Lambda$ and is larger than the possible kinematic 
energy of dark matter of a maximum speed of 600 km/s.  Therefore 
dark matter will not scatter through vector interaction with
Xe nuclei on tree level.

The axial-vector coupling of $\chi_d$ with Z still needs consideration. 
This coupling is elastical and of the size ${\cal O}(\frac{\Delta m}{\mu'})$, 
which is between $10^{-4}$ to $10^{-6}$. It will bring 
to spin-dependent cross section with Xe. 
The spin-dependent tree level cross section for WIMP 
elastically scattering at zero momentum transfer is\cite{sdcross}:
\begin{equation}
\sigma_{SD}=\frac{32(J+1)}{\pi J}G_F^2M_r^2|a_p\langle S_p\rangle 
+ a_n\langle S_n\rangle |^2
\end{equation}
where J is nuclear spin, $M_r$ is the reduced mass of 
dark matter and target nuclear, $a_{p,n}$ are effective proton 
and neutron couplings, $\langle S_{p,n}\rangle$ are 
spin expectation values of proton and neutron in the 
nuclear. The $a_{p,n}$ for $\chi_d$ is of the magnitude 
$\frac{\Delta m}{\mu'}$.  
The $\chi_d$-Xe spin-dependent cross section is 
$(10^{-45}-10^{-41})$ cm$^2$ 
for Xe$^{129}$ and $(10^{-46}-10^{-42})$ cm$^2$ for Xe$^{131}$.
These cross sections are far beyond the detection capability 
of Xenon100 or Xenon 1t.

At one-loop level, the dark matter-nuclear cross section is essentially 
the same as that discussed in Ref. \cite{crosssection}: the one-loop 
$\chi_d$-nucleon spin-independent cross section is about 
$10^{-48}$ cm$^2$, which is also too small to be detected.  

Dark matter in our model cannot produce positron excess observed by 
cosmic-ray experiments such as PAMELA and Fermi-LAT\cite{pamela,fermi}. 
The reason is that there is no light particle to provide 
enough Sommerfeld enhancement for dark matter annihilation in our model. 
As a result, we have to consider the observed positron excess 
as an astrophysical phenomenon.

\subsection{Collider phenomenology}  

Experimental constraints should be considered.  In this model, particles 
beyond those of MSSM, which also carry color or electric charges are 
$D_4^c, D_H^c$ and $\chi_1^-, \chi_2^+$.  Each pair forms an exact Dirac 
particle.  The former is down-type quark like, and the latter charged 
lepton like.  The direct experimental search at LEP requires that they 
should be heavier than 100 GeV, that at Tevatron for down-type heavy 
quarks requires they are heavier than 372 GeV \cite{pdg}, and at LHC 
the current limit is 675 GeV for the down-type heavy quark \cite{lhc}.  
These results can be simply satisfied 
if $\mu^D$ is larger than 675 GeV and $\mu'$ larger than 100 GeV.  
The electroweak precision measurements generally have weak constraints 
on this model, because these extra matters are vector-like which have 
contribution in the form of $1/\mu^{D(\prime)2}$ as expected from 
the decoupling theorem.  The effect of the extra matters can be small 
enough $\leq\left(m_t/\mu^{D(\prime)}\right)^2\simeq (1-10)\%$ if we 
take $\mu^{D(\prime)}\simeq 500$ GeV $-1$ TeV.  Noting that these direct 
search limits are obtained with assumption of single decay channel 
dominant, we will take that $\mu^{D}= 500$ GeV and 
$\mu^{\prime}\simeq 1.1$ TeV in numerical illustration.   
 
There are constraints coming from the unitarity of the $3\times 3$ CKM 
quark mixing matrix of three chiral generations \cite{pdg}.  This 
unitarity is consistent with current data within experimental errors.  
In this model, extra down-type quarks mix with ordinary three chiral 
down-type quarks, which necessarily break the unitarity of the CKM mixing 
matrix.  Unitarity violation is about $(m_{i4}/\mu^D)^2$ where the mass 
parameter $m_{i4} \equiv -y_i^D v\cos\beta/\sqrt{2}$ as seen from Eq. 
(\ref{04}).  This $\mu^D$ dependence is generally expected in the case 
of extra vector-like quarks.  Hierarchical or small mixing masses 
$m_{i4}$ can easily make the CKM matrix approximately unitary within 
errors.  For an example, $(m_{14}/\mu^D)^2\leq 10^{-3}$.  Assuming only 
the third generation mixes with extra quarks, the constraint is still 
loose, $(m_{34}/\mu^D)^2\leq 0.39$.  The quantity $m_{34}$ is at most 
about $m_t$.  This gives that the parameter $\mu^D \geq 280$ GeV.  
It is easy to see that there are new phases in fermion mixing matrices.  
However, these new matrix elements are of order of 
$\left(m_t/\mu^D\right)^2$ at most.  So new CP violation effects are 
generally suppressed.  

Decay signals of these new particles can be easily identified.  From 
trilinear Yukawa interactions given in Eq. (\ref{04}), it is seen that 
$D_4^c, D_H^c$ decay into SM first three generation matters.  Denoting 
the new Dirac quark in $(D_4^c, D_H^c)$ as $\Psi_D$, decays of $\Psi_D$ 
have following results, 
\begin{equation}
\label{37}
\Gamma(\Psi_D\to d_i^c\; \; h^0)\simeq\frac{1}{16\pi}|y_i^D|^2|\mu^D|
\left(1-\frac{m_h^2}{|\mu^D|^2}\right)^2  \;.
\end{equation} 
Taking relevant Yukawa coefficients $y_i^D$'s $\sim~10^{-1}-10^{-2}$, 
the decay rate in Eq. (\ref{37}) is $\Gamma \sim 5-500$ MeV.  Taking 
EWSB into consideration, $\Psi_D$ mixes with SM fermions.  We see that 
the decay $\Psi_D \rightarrow \bar{t}~W^+$ occurs via the 
SU(2)$_L$ gauge interaction at the level of ${\mathcal O}(m/\mu^D)$, 
\begin{equation}
\label{39}
\begin{array}{lll}
\Gamma(\Psi_D \rightarrow \bar{t}~W^+) & \simeq & \displaystyle
\frac{G_F m_W^2|\mu^D||V_{35}|^2}{8\sqrt{2}\pi}
\left\{1+\left(\frac{m_W}{\mu^D}\right)^4
+\left(\frac{m_t}{\mu^D}\right)^4 
-2\left[\left(\frac{m_W}{\mu^D}\right)^2
+\left(\frac{m_W}{\mu^D}\right)^2\left(\frac{m_t}{\mu^D}\right)^2
\right.\right.
\\[3mm]
& & \displaystyle \left.\left. 
+\left(\frac{m_t}{\mu^D}\right)^2\right]\right\}^{1/2}
\left\{\left[1-\left(\frac{m_t}{\mu^D}\right)^2\right]^2 
+\left(\frac{m_W}{\mu^D}\right)^2\left[1+
\left(\frac{m_t}{\mu^D}\right)^2\right]
-2\left(\frac{m_W}{\mu^D}\right)^4
\right\} \; 
\end{array}
\end{equation}
where $V_{35}$ is the mixing element, and the phase space factors 
were given in Refs. \cite{phase}.  Taking $m/\mu^D \sim 1/3$, the 
$\Gamma$ is about 1 GeV.  
In this decay process, the top quark further decays into three quarks 
with one of them a bottom, and the $W$ can decay into a single charged 
lepton with a neutrino.  Taking this process as the main decay channel, 
in the case of $\Psi_D$ pair production, the signal can be searched in 
the events of 2 charged leptons (electron or muon) and 6 jets with two 
of them b-jets, and large missing energies.   

Denoting the heavy charged Dirac lepton ($\chi_1^-, \chi_2^+$) as 
$\Psi_{-}$, its mass is only $\Delta M \sim$ (0.1 - 1) GeV above the 
dark matter.  It can decay into $\Psi_d$ or $\Psi_d'$ together with a 
pion or with a charged lepton (electron or muon) and a neutrino.  In 
the limit of $\Delta M << \mu'$, the typical decay width of $\Psi_{-}$ 
is
\begin{equation}
\label{40}
\Gamma(\Psi_- \rightarrow \Psi_d~e^-~\bar{\nu}_e)  \simeq  
\displaystyle \frac{G_F^2 \Delta M^5}{15\pi^3}\,.
\end{equation}
This lifetime is about $10^{-12} - 10^{-7}$ s.  For a larger $\Delta M$, 
$\Psi_{-}$ decay rapidly into a charged lepton and missing energy, in 
which situation detection is difficult.  For $\Delta M<$ 0.3 GeV, 
$\Psi_{-}$ is long-lived and can leave tracks in detectors.  As $\Psi_-$  
is always produced in pairs, the signal should be easy to identify.

The new quark can be produced at LHC via the gluon fusion process, 
$g~~g \rightarrow \Psi_D \bar{\Psi_D}$.  The production mechanism is 
essentially the same as that of the top quark \cite{bn} with an 
estimated cross section $\sim$ hundreds $fb$ by taking $\mu^D \sim 500$ 
GeV and $\sqrt{s}=14$ TeV.  For the new charged lepton, the Drell-Yan 
process is the main production mechanism.  The cross section is 
estimated to be a few $fb$ which means a few events in one year at most 
\cite{fc}.

\section{discussion}

Finally we will discuss some physical aspects related to this model.  
First of all, it is important to discuss GUT.  We have kept gauge 
coupling unification at the high scale, but a real GUT model with a 
simple gauge group has not been given, it is still far from our reach.  
From the particle content and assumptions of this model, it is doubtful 
if such a GUT model exists.  We wonder if a simple gauge group is really 
necessary for the so-called unification.  In fact, in certain string 
models, unification is achieved without requiring a GUT gauge group, but 
the gauge coupling constants are unified at the scale of $10^{16}$ GeV 
or so \cite{weinberg}.  It would be nice if this model can be 
reconstructed as a string model.  

The assumption of baryon number conservation may have a better looking.  
It has been shown that it can be replaced by that of a $Z_3$ discrete 
symmetry which is called baryon parity \cite{ibanez}.  As we know that 
any global symmetry is not favored from the point of view of the 
quantum gravity, because black holes violate such symmetries.  However, 
the baryon parity can be considered as a result of gauge symmetry 
breaking \cite{discrete}.  

The two U(1) gauge interactions have a mixing.  With such a mixing which 
is symmetry allowed, it is guaranteed that dark matter is composed of 
$\chi_d$ only.  The other particles of the dark sector can decay via the 
mixing even if they are lighter than $\chi_d$.  This is a simple 
scenario of dark matter in this model, although a complicated one with 
vanishingly small $\epsilon$ is possible.  Without the mixing, the 
lightest particle of the dark sector other than $\chi_{1,2}$ would be 
stable and contribute to the relic density of dark matter.  In that 
case, dark matter would be composed of multiple components.  

Our dark matter model has nothing to do with the so-called indirect 
indication of the dark matter by astrophysical observation of 
ATIC \cite{atic}, PAMELA \cite{pamela} and FermiLAT \cite{fermi}.  It 
looks that our model has a potential to accommodate their observation.  
They have observed an access of cosmic electrons and positrons.
Although Fermi-LAT experiment \cite{fermi} does not support ATIC, it 
agrees with PAMELA.  The observation has inspired a lot of theoretical 
re-consideration about the WIMP dark matter \cite{newdm,newdm2,decay}.  
Arkani-Hamed {\it et al.} \cite{newdm} have proposed a scenario to 
understand all experiments on the dark matter in a natural way.   WIMPs 
should have new interaction mediated by a light, GeV scale particle 
which enhances their current annihilation cross section via the 
Sommerfeld mechanism.  This GeV particle is supposed to be the main 
annihilation product, and because of its lightness, it finally decays 
into leptons only.  There are many theoretical models realizing this 
new scenario \cite{newdm2}.  There is a point which is similar to our 
case.  In our model, the WIMP also has new interaction which, however, 
has a typical energy scale of hundreds GeV.  If we had made some tuning 
on the mass parameter $\mu''$ to reduce the new interaction scale to be 
GeV, this model would realize most part of the scenario of Arkani-Hamed 
{\it et al.}.  But because of the dark matter structure of this model, 
the Sommerfeld enhancement factor would be about $10^4$ which has been 
ruled out \cite{feng}.

\section{Summary}

Within the framework of R-parity violation, we have studied the dark 
matter problem with the constraint of gauge coupling constant 
unification.  The WIMP dark matter is contained in a new vector-like 
SU(2)$_L$ doublet which possesses a new U(1) gauge symmetry.  The Higgs 
particles are included in a ${\bf 5 \oplus \bar{5}}$ representation of 
SU(5).  Instead of R-parity, baryon number conservation is assumed.  
In this model, the dark matter particle is stable as a result of the 
gauge invariance.  An accidental discrete $Z_2$ symmetry remains after 
spontaneous gauge symmetry breaking, which makes the dark matter 
particle stable. 

Main results of this model is the following.  
The mass of the dark matter particle is $(1.1 - 1.5)$ TeV.  The dark 
matter and nucleus interaction has small cross sections which are 
consistent with the current dark matter direct detection experiments 
like Xenon100.  In addition to the particle content of MSSM, we have a 
new down-type Dirac quark and a new Dirac charged lepton with masses 
about $(500 - 1000)$ GeV.  These charged particles, especially the new 
quark can be produced at the LHC and hopefully be observed.

\begin{acknowledgments}
We would like to thank Xiao-jun Bi, Yu-feng Zhou and Shou-hua Zhu for 
helpful discussions.  This work was supported in part by the National 
Science Foundation of China under Grant Nos. 11075193, and 10821504, by 
the National Basic Research Program of China under Grant No. 
2010CB833000, and by CAS via KJCX3-SYW-N2.
\end{acknowledgments}


\newpage

\begin{thebibliography}{99}

\bibitem{review1}
For the earliest, see F. Zwicky, Helv. Phys. Acta 6 (1933) 110; \\
For a review, see G. Bertone, D. Hooper and J. Silk, 
Phys. Rept. 405 (2005) 279, and references therein.  

\bibitem{review2}
For reviews, see E.W. Kolb and M.S. Turner, {\it The Early Universe}, 
(Addison-Wesley, Redwood City, CA, 1990);
M. Drees and G. Gerbier, Phys. Lett. B 667 (2008) 241; 
J.L. Feng, J. Phys. G 32 (2006) R1.  

\bibitem{gut}
H. Georgi and S. L. Glashow, Phys. Rev. Lett. 32 (1974) 438; \\
J.C. Pati ans A. Salam, Phys. Rev. D 10 (1974) 275.

\bibitem{susy}
Y. Gol¡¯fand and E. Likhtman, JETP Lett. 13 (1971) 323; 
D.V. Volkov and V. Akulov, Phys. Lett. B 46 (1973) 109; 
J. Wess and B. Zumino, Nucl. Phys. B 70 (1974) 39; 
E. Witten, Nucl. Phys. B 188 (1981) 513; 
S. Dimopoulos and H. Georgi, Nucl. Phys. B 193 (1981) 150; 
N. Sakai, Z. Phys. C 11 (1981) 153; 
L. Ibanez and G.G. Ross, Phys. Lett. B 110 (1982) 215;
L. Alvarez-Gaume, M. Claudson and M.B. Wise, Nucl. Phys. B 207 (1982) 96.

\bibitem{lep}
P. Langacker, M.-X. Luo, Phys. Rev. D 44 (1991) 817;
C. Giunti, C.W. Kim, U.W. Lee, Mod. Phys. Lett. A 6 (1991) 1745;
U. Amaldi, W. de Boer, H. Furstenau, Phys. Lett. B 260 (1991) 447;
J. Ellis, S. Kelley, D. Nanopolous, Phys. Lett. B 260 (1991) 131.

\bibitem{newdm}
N. Arkani-Hamed, D.P. Finkbeiner, T.R. Slatyer and N. Weiner, 
Phys. Rev. D 79 (2009) 015014; \\
N. Arkani-Hamed and N. Weiner, JHEP 0812 (2008) 104.  

\bibitem{1} D.R. Smith and N. Weiner, Phys. Rev. D 64 (2001) 043502; \\  
G. Belanger, A. Pukhov and G. Servant, JCAP 0801 (2008) 009.  

\bibitem{liu} 
C. Liu, Phys. Rev. D 80 (2009) 035004. 

\bibitem{weiner}
D.P. Finkbeiner, L. Goodenough, T.R. Slatyer, M. Vogelsberger 
and N. Weiner, JCAP 1105 (2011) 002

\bibitem{mdm}
M. Cirelli, A. Strumia, New J. Phys. 11 (2009) 105005.  

\bibitem{witten} 
M. W. Goodman, E. Witten, Phys. Rev. D 31, 3059 (1985)

\bibitem{xenon}
E. Aprile {\it et al.} (Xenon100 Collaboration), Phys.Rev.Lett. 107 (2011) 131302 .

\bibitem{sdcross}
J.D. Lewin and P. Smith, Astrop. Phys.6 (1996) 87, 
J. Engel et al., Int. J. Mod. Phys. E1 (1991)

\bibitem{crosssection}
J. Hisano, K. Ishiwata, N. Nagata and T. Takesako, JHEP 1107 (2011) 005.  

\bibitem{pdg} 
J. Beringer {\it et al.} (Particle Data Group), 
Phys. Rev. D 86 (2012) 010001.  

\bibitem{lhc} 
S. Chatrchyan {\it et al.} (CMS Collaboration), arXiv:1210.7471 
[hep-ex];  
For a good review, 
see Y. Okada and L. Panizzi, arXiv:1207.5607 [hep-ph].  

\bibitem{phase}
M. Je\'zabek and J.H. K\"uhn, Nucl. Phys. B 314 (1989)1; \\
J. Reid, G. Tupper, G. Li and M.A. Samuel, Z. Phys. C 51 (1991) 395.  

\bibitem{bn}
E.L. Berger and H. Contopanagos, 
Phys. Lett. B 361 (1995) 115, hep-ph/9512212;\\
P. Nason, S. Dawson and R. K. Ellis, Nucl. Phys. B 303 (1988) 607; \\
For a review, see T. Han, hep-ph/0508097.

\bibitem{fc}
P.H. Frampton, D. Ng, M. Sher and Y. Yuan, Phys. Rev. D 48 (1993) 3128;\\
J.E. Cieza Montalvo and P.P. de Queiroz Filho, 
Phys. Rev. D 66 (2002) 055003; \\
C. Liu and S. Yang, Phys. Rev. D 81 (2010) 093009. 

\bibitem{Belanger:2010pz}
G.~Belanger, F.~Boudjema, A.~Pukhov and A.~Semenov,
arXiv:1005.4133 [hep-ph].

\bibitem{Belanger:2008sj}
G.~Belanger, F.~Boudjema, A.~Pukhov and A.~Semenov,
arXiv:0803.2360 [hep-ph].

\bibitem{Belanger:2006is}
G.~Belanger, F.~Boudjema, A.~Pukhov and A.~Semenov,
Comput.\ Phys.\ Commun.\  {\bf 176} (2007) 367
[arXiv:hep-ph/0607059].  

\bibitem{weinberg}
S. Weinberg, arXiv:hep-ph/9211298.  

\bibitem{ibanez}
L.E. Ibanez and G.G. Ross, Nucl. Phys. B 368 (1992) 3.

\bibitem{discrete}
L.M. Krauss and F. Wilczek, Phys. Rev. Lett. 62 (1989) 1221.

\bibitem{atic} 
J. Chang {\it et al.} (ATIC Collaboration), Nature 456 (2008) 362.  

\bibitem{pamela} 
O. Adriani {\it et al.} (PAMELA Collaboration), Nature 458 (2009) 607.  

\bibitem{fermi} 
A.A. Abdo {\it et al.} (Fermi LAT Collaboration), 
Phys. Rev. Lett. 102, 181101 (2009);
M. Ackermann {\it et al.} (Fermi LAT Collaboration), 
Phys. Rev. Lett. 108, 011103 (2012)  

\bibitem{newdm2}
For reviews, see X.-G. He, Mod. Phys. Lett. A 24 (2009) 2139; \\
J.-M. Yang, Mod. Phys. Lett. A 25 (2010) 976.  

\bibitem{decay}
P.-F. Yin, Q. Yuan, J. Liu, J. Zhang, X.-J. Bi, 
S.-H. Zhu and X.-M. Zhang, Phys. Rev. D 79 (2009) 023512; 
J.T. Ruderman and T. Volansky, JHEP 1002 (2010) 024; 
W.-L. Guo, Y.-L. Wu, Y.-F. Zhou, Phys. Rev. D 81 (2010) 075014; 
M.-X. Luo, L.-C. Wang, W. Wu, and G.-H. Zhu, 
Phys. Lett. B 688 (2010) 216; 
X. Gao, Z. Kang and T. Li, Eur. Phys. J. C 69 (2010) 467. 

\bibitem{feng}
J. Feng and H.-B. Yu, Phys. Rev. D 82 (2010) 083525.  

\end{thebibliography}
\end{document}